\documentclass[11pt]{article}
\usepackage{graphicx}
\usepackage{amsmath}
\usepackage{amsfonts}   
\usepackage{amssymb}

\begin{document}
\date{}
\title{{\bf{\Large Glassy Phase Transition and Stability in Black Holes}}}
\author{
 {\bf {\normalsize Rabin Banerjee}$
$\thanks{e-mail: rabin@bose.res.in}},\, 
 {\bf {\normalsize Sujoy Kumar Modak}$
$\thanks{e-mail: sujoy@bose.res.in}}\\
 {\normalsize S.~N.~Bose National Centre for Basic Sciences,}
\\{\normalsize JD Block, Sector III, Salt Lake, Kolkata-700098, India}
\\[0.3cm]
{\bf {\normalsize Saurav Samanta}$
$\thanks{e-mail: srvsmnt@gmail.com}}\\
 {\normalsize Narasinha Dutt College}
\\{\normalsize 129, Belilious Road, Howrah-711101, India}
\\[0.3cm]
}

\maketitle

\begin{abstract}
Black hole thermodynamics, confined to the semi-classical regime, cannot address the thermodynamic stability of a black hole in flat space. Here we show that inclusion of correction beyond the semi-classical approximation makes a black hole thermodynamically stable. This stability is reached through a phase transition. By using Ehrenfest's scheme we further prove that this is a glassy phase transition with a Prigogine-Defay ratio close to 3. This value is well placed within the desired bound (2 to 5) for a glassy phase transition. Thus our analysis indicates a very close connection between the phase transition phenomena of a black hole and glass forming systems. Finally, we discuss the robustness of our results by considering different normalisations for the correction term.
\end{abstract}

\section{Introduction}
Black holes are the most striking predictions of Einstein's general theory of relativity of gravitation. A possible identification of black holes as thermodynamical systems has unanimously evolved over the years.

The study of black hole thermodynamics is therefore quite important. However, if one analyses properties like thermodynamical stability, phase transition, black hole evaporation etc. the semi-classical results in the canonical ensemble are not much encouraging\cite{BHT}. This is because the specific heat is always negative and therefore in a canonical ensemble it is not possible to attain a stable equilibrium with the surroundings. However, results in a microcanonical ensemble indicate the possibility of stability \cite{BHT,davies}. 


The aim of the present paper is to make a thorough analysis of the possibility of phase transition and stability in black holes, within the framework of a canonical ensemble. As a specific example we consider the Kerr black hole which provides a striking difference from the usual spherically symmetric solutions of Einstein's equation. The main point is the appearance of a work term like $\Omega dJ$, where $\Omega$ and $J$ are the angular velocity and angular momentum, respectively, in the first law of black hole thermodynamics. As shown later, this is the crucial term for classifying phase transition. As we have already hinted in the previous paragraph it is necessary to go beyond the standard semiclassical approximation to discuss phase transition. The results for various thermodynamic variables, for the Kerr black hole, beyond the semiclassical approximation were recently obtained by two of us \cite{Modak}. In particular, the entropy was shown to receive logarithmic corrections. Such corrections have some interesting applications in braneworld cosmology \cite{Lidsey:2002ah}. In this picture the FRW equations get modified leading to a different early/late time cosmology. Also, it has been suggested that \cite{Lidsey:2002ah} in the presence of logarithmic corrections, small black holes can be stable even in a flat back ground. Our calculations in this paper actually quantify such suggestions by systematically analysing phase transition and aspects of stability. Although the logarithmic corrections to be used here were obtained in \cite{Modak} the utility of such terms in the context of phase transition were neither stated nor analyzed in \cite{Modak}. This is the new input in this paper.

Using the corrected expressions of different thermodynamic entities derived in \cite{Modak} we show that there exists a phase transition through which a black hole can switch over to a stable phase from an initial unstable phase. The continuity of the corrected entropy for any value of the black hole mass rules out the existence of a first order phase transition. However we find that there is a discontinuity in specific heat for a critical value of the mass. At this mass the black hole temperature is maximum. Then it sharply decreases finally attaining the absolute zero for a vanishing mass. Therefore this indicates a transition from an unstable phase to a stable phase. Having found this possibility of phase transition, we then analyze the nature of this phase transition. We first derive two Ehrenfest like equations for the rotating Kerr black hole. If this phase transition is second order in nature the Ehrenfest's equations must be satisfied. A numerical analysis reveals that the first Ehrenfest's equation is satisfied for a long range of angular velocity but the second Ehrenfest's equation is violated. Hence a second order phase transition is also ruled out.

The final possibility is the occurrence of a glassy phase transition. It is found from experiments that in a liquid to glass phase transition, the specific heat ($C_p$), compressibility ($k$) and thermal expansivity ($\alpha$) {\it i.e.} the second order derivatives of the Gibbs free energy, make a smeared jump. For this type of transition the Prigogine-Defay (PD) ratio \cite{jackle}
measures the deviation from the second Ehrenfest's equation. For a second order phase transition $\Pi=1$ exactly but for a glassy phase transition $\Pi$ varies from 2 to 5.

In the present paper we have shown that the second derivatives of free energy suffer a smeared discontinuity at the phase transition point and the PD ratio as found by us for the Kerr black hole is nearly 3. This enables us to conclude that the transition of the black hole from an unstable to a stable phase is a glassy phase transition.

In this paper the coefficient of the correction term to the semi-classical entropy is taken to be $\frac{1}{90}$ which was obtained in \cite{Modak}. To check the robustness of our conclusion it is desirable to take other normalisations. However, here we are faced with a lack of data. This is because, apart from our work \cite{Modak}, there is no other computation which explicitly gives the coefficient of the logarithmic term for the Kerr black hole. We therefore choose arbitrarily some values for this normalisation. Specifically we have taken this factor to be $\frac{1}{2}$ (which is also less than 1, like $\frac{1}{90}$) and $\frac{3}{2}$ (which is greater than 1) as well as 1. In all cases we observe a glassy phase transition with the same PD ratio. This establishes the soundness of our conclusion.

The novelty of this paper is not only confined to its results but also to the approach. The issues related to the thermodynamic stability of a black hole have never been explored by going beyond the semi-classical regime. As we have shown it is essential to go beyond this approximation to study thermodynamic stability through a phase transition. Of course there is a wide literature on the Hawking-Page \cite{Hawkpage} phase transition which however occurs only in a black hole of AdS space \cite{group}. In this paper, on the other hand we discuss phase transition and thermodynamic stability in an asymptotically flat Kerr black hole.  
  
The paper is organised in the following manner. In section 2 we study the Kerr black hole as a thermodynamical system, staying within the semi-classical regime. In the next section we go beyond the semi-classical approximation and discover a phase transition for the Kerr black hole. Section 4 is devoted to analyse the nature of this phase transition. Here we show that it is a glassy phase transition which makes the Kerr black hole thermodynamically stable. We give our conclusions in section 5. There are two appendices presented at the end of the paper. In appendix 1 we derive two Ehrenfest's equations considering the Kerr black hole as a thermodynamical system and in appendix 2 basic results are compiled to prove the robustness of our conclusions for different normalisations of the correction (logarithmic) term.

\section{Kerr black hole as a thermodynamical system; a semi-classical study}
The Kerr black hole is an axisymmetric, non-static solution of the Einstein equation having two Killing vectors. It has two conserved parameters mass $(M)$ and angular momentum $(J)$ corresponding to those vectors. The location of outer/event ($r_+$) and inner/Cauchy ($r_-$) horizons are given by   
\begin{equation}
r_{\pm}=M\pm \sqrt{M^{2}-a^{2}},
\label{equation 4}
\end{equation}
where 
\begin{equation}
a=\frac{J}{M}~~(a\neq0)
\label{equation 4a}
\end{equation}
is the angular momentum per unit mass. The extremal condition (where the inner and the outer horizon coincide) for the Kerr black hole is given by $M^2=\frac{J^2}{M^2}$. The angular velocity and area of the event horizon are defined as \cite{Carrol}
\begin{eqnarray}
\Omega_{H}= \frac{a}{r^2_+ + a^2}.
\label{equation 5}
\end{eqnarray}
and
\begin{eqnarray}
A= 4\pi (r^2_+ + a^2)
\label{equation 6}
\end{eqnarray}
respectively. The semi-classical Hawking temperature and entropy of the rotating Kerr black hole are given by \cite{Carrol}
\begin{eqnarray}
T=\frac{\hbar\kappa}{2\pi}=\frac{\hbar}{2\pi}\frac{\sqrt{M^2-a^2}}{(r^2_+ +a^2)}.
\label{equation 7}
\end{eqnarray}
and 
\begin{equation}
S=\frac{A}{4\hbar}=\frac{\pi(r_+^2+a^2)}{\hbar}.
\label{equation 8}
\end{equation}
respectively. Note that for the extremal Kerr black hole the semi-classical Hawking temperature vanishes. To calculate the semi-classical specific heat it is desirable to first make the appropriate identifications. Comparison between the work terms of the first law of thermodynamics and first law of black hole thermodynamics, 
\begin{eqnarray}
dE= TdS - PdV,\label{equation 2}\\
dM=TdS+\Omega_H dJ,\label{equation 3a}
\end{eqnarray}
yields the following map  
\begin{eqnarray}
P \rightarrow -\Omega_H,~~V\rightarrow J. 
\label{equation 9}
\end{eqnarray}
This is a well known analogy where one can treat $-\Omega_H dJ$ as the external work done on the black hole \cite{Carrol}.

The semi-classical specific heat for the Kerr black hole can now be written, following its standard thermodynamic definition, as, 
\begin{eqnarray}
C_{\Omega_H}=T\left(\frac{dS}{dT}\right)_{\Omega_H}=\left(\frac{dM}{dT}\right)_{\Omega_H}-\Omega_{H}\left(\frac{dJ}{dT}\right)_{\Omega_H}
\label{equation 10}
\end{eqnarray}
where, in the second equality, (\ref{equation 3a}) has been used. For convenience, instead of working with the variables ($M,\Omega_H,J$), we take ($M,\Omega_H,a$) as the basic variables and express $\left(\frac{dJ}{dT}\right)_{\Omega_H}$ as  
\begin{eqnarray}
\left(\frac{dJ}{dT}\right)_{\Omega_H}=a\left(\frac{dM}{dT}\right)_{\Omega_H}+M\left(\frac{da}{dT}\right)_{\Omega_H}
\label{equation 11}
\end{eqnarray}
where (\ref{equation 4a}) has been used. From (\ref{equation 10}) and (\ref{equation 11}) we find 
\begin{eqnarray}
C_{\Omega_H}=(1-a\Omega_H)\left(\frac{dM}{dT}\right)_{\Omega_H}-M\Omega_{H}\left(\frac{da}{dT}\right)_{\Omega_H}.
\label{equation 12}
\end{eqnarray}
To calculate $C_{\Omega_H}$, we first recast the semi-classical Hawking temperature (\ref{equation 7}), by eliminating $r_+$ using (\ref{equation 5}),
\begin{eqnarray}
T=\frac{\hbar\Omega_H\sqrt{M^2-a^2}}{2\pi a}=T(M, \Omega_H, a)
\label{equation 13}
\end{eqnarray}
Likewise from (\ref{equation 4}) and (\ref{equation 5}), we find,
\begin{eqnarray}
a=\frac{4M^2\Omega_H}{4M^2\Omega_H^2+1}\label{equation 14}
\end{eqnarray}
Exploiting this, one of the variables ($M,a$) can be replaced in favour of the others and thus (\ref{equation 13}) can be written in either of the following ways {\footnote{From now on by $a,T,M$ and $\Omega_H$ we mean $\frac{a}{\sqrt\hbar},\frac{T}{\sqrt\hbar},\frac{M}{\sqrt\hbar}$ and $\sqrt\hbar\Omega_H$ respectively. With this convention, $\hbar$ no longer appears explicitly in any of the equations.}},  
\begin{eqnarray}
T=T(M,\Omega_H)=\frac{1}{8\pi M}(1-4M^2\Omega_H^2),\label{equation 16}\\
T=T(\Omega_H,a)=\frac{(1-2a\Omega_H)}{4\pi}\sqrt\frac{\Omega_H}{a(1-a\Omega_H)}.\label{equation 17}
\end{eqnarray} 
Now using these two equations together with (\ref{equation 14}) we find the semi-classical specific heat (\ref{equation 12}) as
\begin{eqnarray}
C_{\Omega_H}=C_{\Omega_H}(M,\Omega_H)= \frac{-8\pi M^2(1-4M^2\Omega_H^2)}{(1+4M^2\Omega_H^2)^3}.
\label{equation 18}
\end{eqnarray}
In the above equation we took $M$ and $\Omega_H$ as the two independent variables for the purpose of graphical analysis. We shall stick to this convention for other cases also. Using (\ref{equation 4}) and (\ref{equation 14}), the semi-classical entropy (\ref{equation 8}) can be expressed in terms of $M, \Omega_H$ as 
\begin{eqnarray}
S=2\pi M\left(M+\sqrt{\frac{M^2(1-4M^2\Omega_H^2)^2}{(1+4M^2\Omega_H^2)^2}}\right).
\label{equation 18a}
\end{eqnarray}
For the non-extremal region ($T>0$), it follows from (\ref{equation 16}) that $(1-4M^2\Omega_H^2)>0$ and therefore one has the following expression for the semi-classical entropy for Kerr black hole,
\begin{eqnarray}
S=S(M,\Omega_H)= \frac{4\pi M^2}{(1+4M^2\Omega_H^2)}.
\label{equation 19}
\end{eqnarray}
Following the thermodynamical definition Gibbs free energy, for the Kerr black hole is defined as,
\begin{eqnarray}
G=M-\Omega_HJ-TS.
\label{equation 19a}
\end{eqnarray}
For Kerr black hole it is found to be,
\begin{eqnarray}
G=\frac{M}{1+4M^2\Omega_H^2}-\frac{1}{4}(1-4M^2\Omega_H^2)\left(M+\sqrt{\frac{(M-4M^3\Omega_H^2)^2}{(1+4M^2\Omega_H^2)^2}}\right).
\label{equation 19aa}
\end{eqnarray}

\begin{figure}[ht]
\centering
\includegraphics[angle=0,width=15cm,height=12cm]{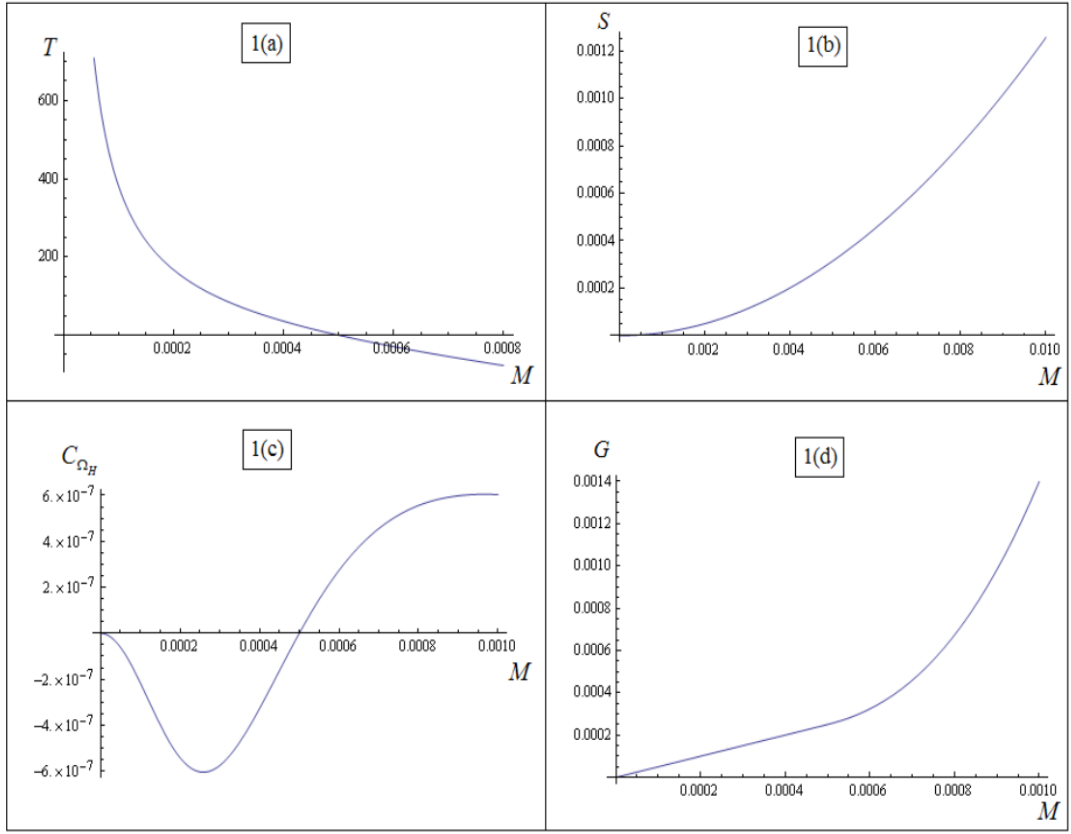}
\caption[]{\it{Semi-classical Hawking temperature ($T$), entropy ($S$), specific heat ($C_{\Omega_H}$) and Gibbs free energy ($G$) vs. mass ($M$) plot: In all curves $\Omega_H=10^3$.}}
\label{figure 1}
\end{figure}
In figure 1 we plot the semi-classical Hawking temperature (\ref{equation 16}), entropy (\ref{equation 18a}), specific heat (\ref{equation 18}) and Gibb's free energy (\ref{equation 19a}) with respect to $M$, keeping the other parameter $\Omega_H$ fixed at a particular value $\Omega_H=10^3$. The temperature diverges as the mass of the black hole tends to zero. The point at which $T=0$ is the extremal case ($M^2=\frac{J^2}{M^2}$) where the Hawking temperature vanishes. From (\ref{equation 16}) this yields $M^2=\frac{1}{4\Omega_H^2}=25\times10^{-8}$. Hence $M=0.0005$ which is also reflected in the $T-M$ graph (figure 1(a)). Beyond this point, the temperature acquires a negative value ($M^2\leq \frac{J^2}{M^2}$) and the curve does not have any physical significance. The entire non-extremal region where $T$ is positive, corresponding to $M^2\geq \frac{J^2}{M^2}$, is relevant. In this non-extremal region specific heat is negative (figure 1(c)) for all values of the mass and thus renders the system unstable. The semi-classical entropy (figure 1(b)) and Gibbs free energy (figure 1(d)) are continuous for the entire non-extremal region. The absence of discontinuity in $S$ and $C_{\Omega_H}$ suggests that there is no phase transition.  Thus at the semi-classical level, the system remains unstable all throughout. Later we shall repeat this analysis by considering correction to the semi-classical Hawking temperature and entropy of the Kerr black hole which will reveal some dramatic changes.


\section{Kerr black hole as a thermodynamical system; beyond semi-classical approximation}

Now we want to investigate the situation when a correction to the semi-classical approximation is taken into account. In a recent work involving two of us \cite{Modak} the black hole temperature and entropy for Kerr black hole were found to be 
\begin{eqnarray}
\tilde T=T\left(1+\frac{\sqrt\hbar}{180\pi Mr_+}+{\cal O}(\hbar^2)\right)^{-1},
\label{equation 20}
\end{eqnarray}    
and
\begin{eqnarray}
\tilde S=\frac{A}{4\hbar}+\frac{1}{90}\log{\frac{A}{4\hbar}}+{\cal O}(\hbar), 
\label{equation 21}
\end{eqnarray} 
where the usual semi-classical expressions for $A$ and $T$ were given in (\ref{equation 6}) and (\ref{equation 7}) respectively. We remark that the WKB type methods adopted in \cite{Modak} were also successfully employed \cite{Modak1} in other black hole geometries reproducing well known results found by other approaches \cite{OTHER}.

 With these structures, the first law of black hole thermodynamics for the Kerr black hole is now given by \cite{Modak}
\begin{eqnarray}
dM= \tilde T d\tilde S+\Omega_HdJ.
\label{equation 24}
\end{eqnarray}
In the literature there exist several approaches which find the logarithmic correction to the semiclassical Bekenstein-Hawking entropy for various black holes\cite{OTHER}. However, the coefficient of the correction term is not identical in these approaches. Also as far as we are aware, there exists no other paper apart from \cite{Modak} which fixes the coefficient of the logarithmic correction to the semiclassical entropy for the Kerr black hole.

In order to calculate the specific heat we follow the method discussed in the previous section and write $\tilde T$ as a function of ($M,\Omega_H$) and ($a,\Omega_H$) in the following way:
\begin{eqnarray}
\tilde T=\tilde T(M,\Omega_H)=T(M,\Omega_H)\left(1+\frac{1+4M^2\Omega_H^2}{360\pi M^2}\right)^{-1},
\label{equation 22a}\\
\tilde T=\tilde T(a,\Omega_H)=T(a,\Omega_H)\left(1+\frac{\Omega_H}{90\pi a}\right)^{-1},
\label{equation 22b}
\end{eqnarray}
where $T$ is just the semi-classical expression defined in (\ref{equation 16}) and (\ref{equation 17}). Using (\ref{equation 14}), the corrected entropy (\ref{equation 21}) can be written as a function of $M,\Omega_H$ in the following way 
\begin{eqnarray}
\tilde S=S+\frac{1}{90}\log{S} 
\label{equation 23}
\end{eqnarray} 
where the semiclassical entropy $S$ is defined in (\ref{equation 18a}). Following the semi-classical analysis, the specific heat is defined as
\begin{eqnarray}
\tilde C_{\Omega_H}=\tilde T\left(\frac{d\tilde S}{d\tilde T}\right)_{\Omega_H}=(1-a\Omega_H)\left(\frac{dM}{d\tilde T}\right)_{\Omega_H}-M\Omega_{H}\left(\frac{da}{d\tilde T}\right)_{\Omega_H}.
\label{equation 25a}
\end{eqnarray}
It is now straightforward to derive the corrected specific heat for the Kerr black hole by using (\ref{equation 22a}), (\ref{equation 22b}) and (\ref{equation 25a}). This finally leads to 
\begin{eqnarray}
\tilde C_{\Omega_H}=-\frac{(1+4M^2\Omega_H^2+360\pi M^2)^2(1-4M^2\Omega_H^2)}{45(1+4M^2\Omega_H^2)^2(16M^4\Omega_H^4+1440\pi M^4\Omega_H^2+16M^2\Omega_H^2+360\pi M^2-1)}.
\label{equation 25b}
\end{eqnarray}
Gibbs free energy, in the presence of quantum correction, is found to be
\begin{eqnarray}
&& \tilde G = \frac{M}{1+4M^2\Omega_H^2}+\nonumber\\
&& \frac{M(4M^2\Omega_H^2-1)\left(180\pi M(M+\sqrt{\frac{(M-4M^3\Omega_H^2)^2}{(1+4M^2\Omega_H^2)^2}})\right)+\log\left(2M\pi(M+\sqrt{\frac{(M-4M^3\Omega_H^2)^2}{(1+4M^2\Omega_H^2)^2}})\right)}{2+8M^2(\Omega_H^2+90\pi)}~~~  
\label{equation 23a}
\end{eqnarray}
Making use of equations (\ref{equation 22a}), (\ref{equation 23}) and (\ref{equation 25b}), we plot $\tilde T$, $\tilde S$, $\tilde C_{\Omega_H}$ and $\tilde G$ with  mass $M$ for the previously fixed value of $\Omega_H=10^3$ (figure 2). Let us discuss the physical significance of the different plots one by one.

\begin{figure}[h]
\centering
\includegraphics[angle=0,width=15cm,height=12cm]{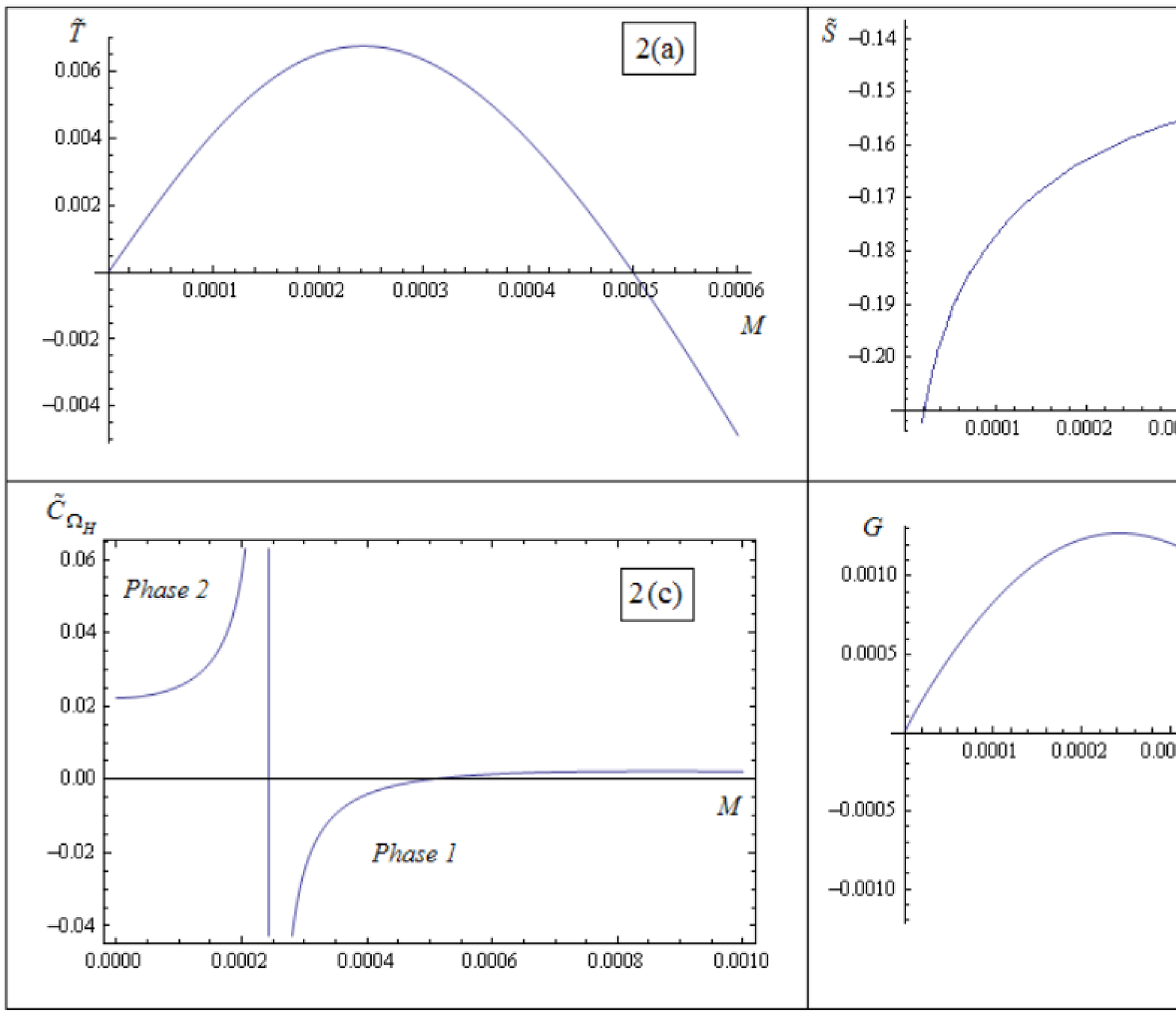}
\caption[]{\it{Corrected Hawking temperature ($\tilde T$), entropy ($\tilde S$), specific heat ($\tilde C_{\Omega_H}$) and Gibbs free energy ($\tilde G$) vs. mass ($M$) plot: In all curves $\Omega_H=10^3$.}}
\label{figure 2}
\end{figure}
Figure 2(a) shows that the corrected Hawking temperature is not diverging when the mass of the black hole goes to zero, rather it vanishes. This behaviour was absent in the semi-classical case. The extremality condition is still satisfied for $M=0.0005$ since at this point the corrected temperature vanishes. There is a critical mass ($M_{\textrm{crit}}=0.0002425$) for which the temperature is maximum given by $T_{\textrm{max}}=T_{\textrm{crit}}=0.006756$. From figure 2(c) one can see that specific heat $C_{\Omega_H}$ now suffers a discontinuity at $M_{\textrm{crit}}$. At this point $\tilde C_{\Omega_H}$ becomes positive from the initial negative value. This point can be elaborated in the following manner. During the black hole evaporation as the mass decreases temperature of the black hole initially increases. This is shown in the region phase 1 of figure 2(c). At $M=M_{\textrm{crit}}$ a phase transition is taking place where the specific heat switches to a positive value. Consequently the black hole cools down during further evaporation and finally at zero mass the temperature vanishes. This behaviour is found in phase 2 region of the graph 2(c). The corrected entropy ($\tilde S$) in figure 2(b) and corrected free energy ($\tilde G$) in figure 2(d), are continuous for the entire region.  

Note that we cannot write the corrected temperature (\ref{equation 20}) as $\tilde T=T\left(1-\frac{\sqrt\hbar}{180\pi Mr_+}+{\cal O}(\hbar^2)\right)$ since at the phase transition point, $\frac{\sqrt \hbar}{180\pi Mr_+}\sim 10^5$. As this is considerably greater than one, a naive simplification of (\ref{equation 20}) by series expansion is not possible.

So far it seems reasonable from the plots that there is a phase transition which eventually leads to the thermodynamic stability of the Kerr black hole. In the next section we shall make a systematic study to identify the type of phase transition. 

\section{Nature of the black hole phase transition}
In the previous section we found a signature of phase transition with the corrected versions of Hawking temperature and entropy of Kerr black hole. This phase transition is not a first order phase transition as there is no discontinuity in the corrected entropy for the entire physical region. Instead, the discontinuity in specific heat suggests that there may be a second order phase transition. If that is the case then one naturally expects that there must be some analog of Ehrenfest's equations in black hole thermodynamics which must be satisfied. Till now Ehrenfest's relations were not studied in the context of black hole thermodynamics. Hence we shall systematically derive (see appendix) the Ehrenfest's equations and then check the validity of these relations afterwards. These (first and second) Ehrenfest's equations are given by (\ref{eren1}) and (\ref{eren2}) respectively. It is noteworthy that, since in a second order phase transition both $S$ and $J$ remain constant, left hand sides of these two equations have the same value. In the next subsection we shall check the validity of these two equations in the context of black hole thermodynamics.

Let us now convince ourselves that there is indeed a discontinuity present in the behaviour of corrected specific heat $C_{\Omega_H}$ (\ref{corsp}), analog of volume expansion coefficient $\alpha$ (\ref{volexp}) and compressibility $k_{T}$ (\ref{compr}). As far as $C_{\Omega_H}$ is concerned, its values at two different phases are given by $C_{\Omega_{H_1}}=0$ and $C_{\Omega_{H_2}}=0.0222$, so that $C_{\Omega_{H_2}}-C_{\Omega_{H_1}}=0.0222$ (figure 2(c)). For others, the expressions for $\alpha$ and $k_T$ in terms of $M$ and $\Omega_H$ are derived and appropriate graphs are plotted.

Comparing (\ref{volexp}) and (\ref{equation 11}) we find
\begin{equation}
J\alpha=a\left(\frac{\partial M}{\partial T}\right)_{\Omega_H}+M\left(\frac{\partial a}{\partial T}\right)_{\Omega_H}
\label{equation 69}
\end{equation}
Using the expressions of the corrected Hawking temperature from (\ref{equation 22a}) and (\ref{equation 22b}) and substituting $a$ from (\ref{equation 14}) this equation is simplified to
\begin{equation}
J\alpha=-\frac{4M^2\Omega_H(1+4M^2(90\pi+\Omega_H^2))^2(4M^2\Omega_H^2+3)}{45(1+4M^2\Omega_H^2)^2[16M^4(90\pi\Omega_H^2+\Omega_H^4)+8M^2(2\Omega_H^2+45\pi)-1]}.
\label{equation 57}
\end{equation}
The plot of $J\alpha$ with the mass ($M$) for our chosen constant angular velocity $\Omega_H=10^3$ of Kerr black hole is shown in figure 3a.
\begin{figure}[ht]
\centering
\includegraphics[angle=0, width=15cm, height=7cm]{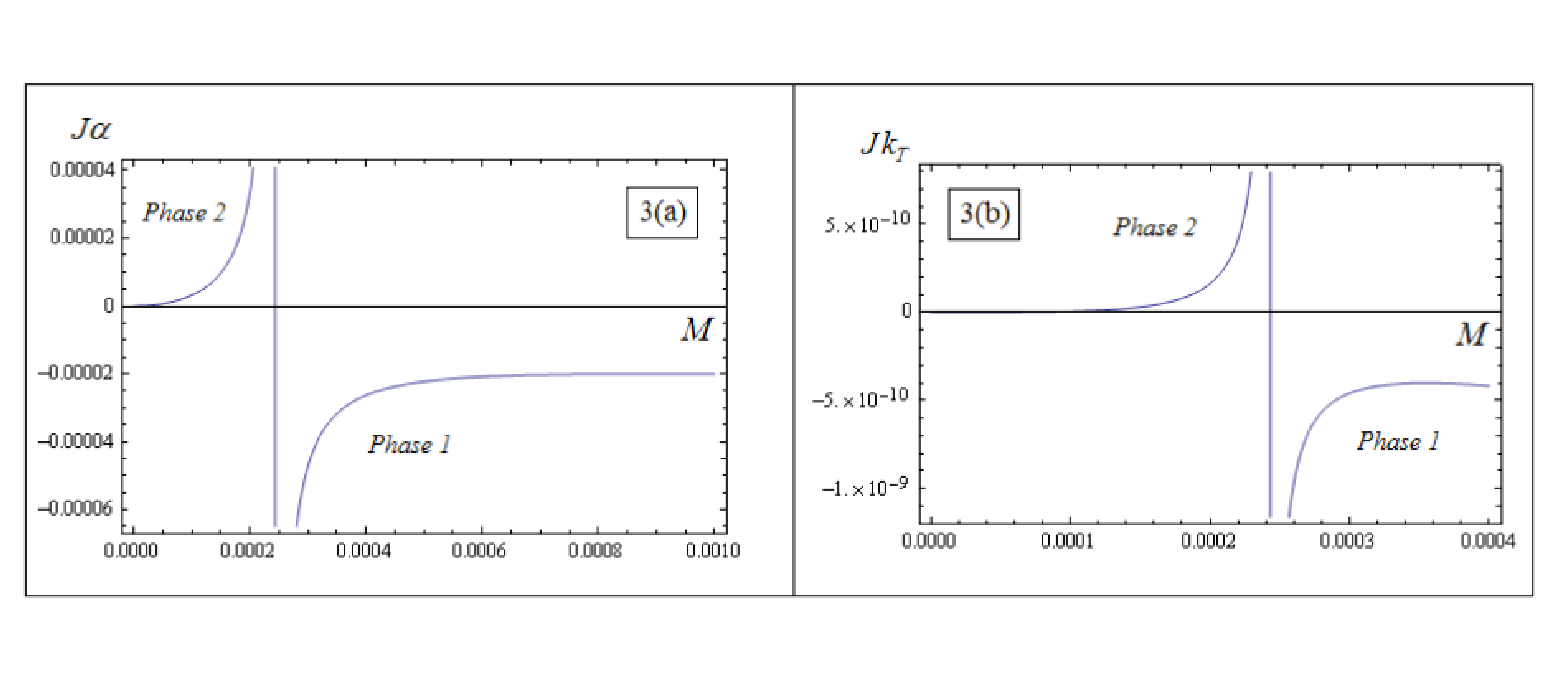}
\caption[]{\it{$J\alpha$, $Jk_T$ vs. mass ($M$) plot for $\Omega_H=10^3$.}}
\label{figure 3}
\end{figure}
 There is a discontinuity present in this plot exactly at the same mass value where the specific heat also suffered a discontinuity. The values of $J\alpha$ in different phases are given by $J{\alpha_1}=-0.000022$ and $J{\alpha_2}=0$.

To express $k_{T}$ in terms of $M$ and $\Omega_H$ we proceed as follows. Using (\ref{equation 4a}) we re-express (\ref{compr}) in the following way, 
\begin{eqnarray}
Jk_T=\left(\frac{\partial J}{\partial\Omega_H}\right)_{\tilde T}=\left(\frac{\partial (Ma)}{\partial\Omega_H}\right)_{\tilde T}\nonumber\\
=M\left(\frac{\partial a}{\partial \Omega_H}\right)_{\tilde T}+a\left(\frac{\partial M}{\partial \Omega_H}\right)_{\tilde T}.
\label{equation 63}
\end{eqnarray}
Using the property of partial differentiation, 
\begin{eqnarray}
\left(\frac{\partial a}{\partial\Omega_H}\right)_{\tilde T}\left(\frac{\partial \Omega_H}{\partial\tilde T}\right)_{a}\left(\frac{\partial\tilde T}{\partial a}\right)_{\Omega_H}=-1;~~\left(\frac{\partial M}{\partial\Omega_H}\right)_{\tilde T}\left(\frac{\partial \Omega_H}{\partial\tilde T}\right)_{M}\left(\frac{\partial\tilde T}{\partial M}\right)_{\Omega_H}=-1
\label{equation 63a}
\end{eqnarray}
we obtain the following two equalities 
\begin{eqnarray}
\left(\frac{\partial a}{\partial\Omega_H}\right)_{\tilde T}=-\frac{\left(\frac{\partial \tilde T}{\partial\Omega_H}\right)_{a}}{\left(\frac{\partial\tilde T}{\partial a}\right)_{\Omega_H}};~~\left(\frac{\partial M}{\partial\Omega_H}\right)_{\tilde T}=-\frac{\left(\frac{\partial\tilde T}{\partial\Omega_H}\right)_{M}}{\left(\frac{\partial\tilde T}{\partial M}\right)_{\Omega_H}}.
\label{equationn 64}
\end{eqnarray}
These two equations can easily be simplified by using (\ref{equation 22a}) and (\ref{equation 22b}). Substituting their values in (\ref{equation 63}) and using (\ref{equation 14}) to eliminate $a$, we finally obtain,
\begin{equation}
Jk_{T}=-\frac{4[M^3+4M^5(7\Omega_H^2-90\pi)+16M^7(540\pi\Omega_H^2+7\Omega_H^4)+64M^9(270\pi\Omega_H^4+\Omega_H^6)]}{(1+4M^2\Omega_H^2)^2[16M^4(\Omega_H^4+90\pi\Omega_H^2)+8M^2(2\Omega_H^2+45\pi)-1]}.
\label{equation 66}
\end{equation}
This is plotted in figure 3(b) for constant $\Omega_H=10^3$. The curve shows a discontinuity at the same critical mass value ($M_{{\textrm{crit}}}=0.0002425$). This plot gives $Jk_{T_{1}}=-4.1\times10^{-10}$ and $Jk_{T_2}=0$. 

The above analysis strongly suggests the occurrence of phase transition. All the relevant physical parameters ($C_{\Omega_H}$, $J\alpha$ and $Jk_{T}$) appearing in the Ehrenfest's equations show discontinuity at the same critical value of mass. To conclusively argue in favour of a phase transition, we explicitly check the validity of the Ehrenfest's equations. In the next subsection we shall address this issue.


\subsection{Validity of Ehrenfest's equations}

Let us consider the first Ehrenfest's equation (\ref{eren1}). Using the chain rule of partial differentiation we have {\footnote {Under the change of variables $u=u(x,y)$ and $v=v(x,y)$ for a function $f=f(x,y)$, $\left(\frac{\partial f}{\partial x}\right)_y=\left(\frac{\partial f}{\partial u}\right)_v\left(\frac{\partial u}{\partial x}\right)_y+\left(\frac{\partial f}{\partial v}\right)_u\left(\frac{\partial v}{\partial x}\right)_y$. In the special case when $u=x$, $\left(\frac{\partial f}{\partial x}\right)_y=\left(\frac{\partial f}{\partial u}\right)_v+\left(\frac{\partial f}{\partial v}\right)_u\left(\frac{\partial v}{\partial x}\right)_y$.}}
\begin{equation}
\left(\frac{\partial\tilde T}{\partial\Omega_H}\right)_{\tilde S}=\left(\frac{\partial {\tilde T}}{\partial \Omega_H}\right)_{a}+\left(\frac{\partial {\tilde T}}{\partial a}\right)_{\Omega_H}\left(\frac{\partial a}{\partial\Omega_{H}}\right)_{\tilde S}
\label{equation 53} 
\end{equation}
The derivatives of the corrected temperature (r.h.s) are computed from (\ref{equation 22b}). To calculate $\left(\frac{\partial a}{\partial\Omega_H}\right)_{\tilde S}$, we first derive an expression for the corrected entropy. On using (\ref{equation 5}) and (\ref{equation 23}) this is written as,
\begin{equation}
\tilde S=\tilde S(a,\Omega_H)=\frac{\pi a}{\Omega_H}+\frac{1}{90}\log{\frac{\pi a}{\Omega_H}}.
\label{equation 54}
\end{equation}
Exploiting the rule of partial differentiation, one obtains    
\begin{eqnarray}
\left(\frac{\partial a}{\partial \Omega_H}\right)_{\tilde S}=-\frac{\left(\frac{\partial{\tilde S}}{\partial\Omega_H}\right)_{a}}{\left(\frac{\partial {\tilde S}}{\partial a}\right)_{\Omega_H}}.
\label{equation 54a}
\end{eqnarray}
From (\ref{equation 54}) and (\ref{equation 54a}) we find $\left(\frac{\partial a}{\partial\Omega_{H}}\right)_{\tilde S}=\frac{a}{\Omega_H}$. Substituting this in (\ref{equation 53}) and using (\ref{equation 22b}) yields,
\begin{equation}
\left(\frac{\partial\Omega_H}{\partial\tilde T}\right)_{\tilde S}=\frac{2(90\pi a+\Omega_H)(1-a\Omega_H)^2}{45a\Omega_H(2a\Omega_H-3)}\sqrt{\frac{\Omega_H}{a(1-a\Omega_H)}}.
\label{equation 55}
\end{equation}
This is the left hand side of the first Ehrenfest's equation. Eliminating $a$ in terms of $M,\Omega_H$ by using (\ref{equation 14}), we find the cherished expression, 
\begin{equation}
-\left(\frac{\partial\Omega_H}{\partial\tilde T}\right)_{\tilde S}=\frac{(1+4M^2\Omega_H^2)(1+4M^2(\Omega_H^2+90\pi))}{180M^3\Omega_H(16M^4\Omega_H^4+16M^2\Omega_H^2+3)}.
\label{equation 55a}
\end{equation}
 For $\Omega_H=10^3$ and $M=M_{\textrm{crit}}=0.0002425$, we get the value of the left hand side of the first Ehrenfest's equation (\ref{eren1}) that follows directly from (\ref{equation 55a}),
\begin{equation}
-\left(\frac{\partial\Omega_H}{\partial\tilde T}\right)_{\tilde S}=1.487\times10^5
\label{equation 58a}
\end{equation}

The right hand side of (\ref{eren1}) is now calculated by using results dictated earlier from different graphs ({\it i.e.} figure 2(a), 2(c), 3(a)). From figure 2(a) we find, at $M=M_{\textrm{crit}}$ the critical temperature $\tilde T=T_{\textrm{crit}}=0.006756$. Using these values in the right hand side of (\ref{eren1}), we get
\begin{eqnarray}
\frac{{\tilde C}_{\Omega_{H_2}}-{\tilde C}_{\Omega{_{H_1}}}}{{\tilde T}_{\textrm{crit}}(J\alpha_2-J\alpha_1)}=\frac{(0.222-0)}{0.006756\times0.000022}=1.494\times10^5
\label{equation 58}
\end{eqnarray}

The remarkable agreement between (\ref{equation 58a}) and (\ref{equation 58}) clearly shows the validity of the first Ehrenfest's equation for the Kerr black hole.

Let us now take the second Ehrenfest's equation (\ref{eren2}). The left hand side of (\ref{eren2}) can be calculated in the following manner. First we simplify (\ref{equation 14}) to find $(1-a\Omega_H)=\frac{J}{4M^3\Omega_H}$. Substituting this in (\ref{equation 22b}) and replacing $a$ from (\ref{equation 4a}) we write the angular momentum in the following way 
\begin{equation}
J=J(M,\Omega_H,\tilde T)=\frac{M\Omega_H(45M-\tilde T)}{90(\pi \tilde T+M\Omega_H^2)}.
\label{equation 58b}
\end{equation}
For constant value of $J$ ({\it i.e.} $dJ=0$) we have
\begin{equation}
\left(\frac{\partial J}{\partial M}\right)_{\Omega_H,\tilde T}dM+\left(\frac{\partial J}{\partial\Omega_H}\right)_{M,\tilde T}d\Omega_H+\left(\frac{\partial J}{\partial \tilde T}\right)_{M,\Omega_H}d\tilde T=0.
\label{equation 59}
\end{equation}
Since $\tilde T$, $M$ and $\Omega_H$ are connected by (\ref{equation 22a}), we take only the variables $M$ and $\Omega_H$ as independent. Then $d\tilde T$ can be written as
\begin{equation}
d\tilde T=\left(\frac{\partial\tilde T}{\partial M}\right)_{\Omega_H}dM+\left(\frac{\partial\tilde T}{\partial\Omega_H}\right)_{M}d\Omega_H.
\label{equation 60}
\end{equation}
Now multiplying (\ref{equation 59}) by $\left(\frac{\partial\tilde T}{\partial M}\right)_{\Omega_H}$ and (\ref{equation 60}) by $\left(\frac{\partial J}{\partial M}\right)_{\Omega_H,\tilde T}$ and subtracting one from the other we arrive at the result
\begin{equation}
\left(\frac{\partial\Omega_H}{\partial\tilde T}\right)_J=\frac{\left(\frac{\partial J}{\partial\tilde T}\right)_{M,\Omega_H}\left(\frac{\partial\tilde T}{\partial M}\right)_{\Omega_H}+\left(\frac{\partial J}{\partial M}\right)_{\tilde T,\Omega_H}}{\left(\frac{\partial J}{\partial M}\right)_{\Omega_H,\tilde T}\left(\frac{\partial\tilde T}{\partial\Omega_H}\right)_{M}-\left(\frac{\partial J}{\partial\Omega_H}\right)_{M,\tilde T}\left(\frac{\partial\tilde T}{\partial M}\right)_{\Omega_H}}
\label{equation 61}
\end{equation}
Derivatives of $J$ and $\tilde T$, obtained from (\ref{equation 58b}) and (\ref{equation 22a}) respectively, are substituted in (\ref{equation 61}) to finally yield the left hand side of the second Ehrenfest's equation (\ref{eren2}) as
\begin{equation}
-\left(\frac{\partial\Omega_H}{\partial\tilde T}\right)_J=\frac{\Omega_H(4M^2\Omega_H^2+3)(1+4M^2\Omega_H^2+360M^2\Omega_H^2)^2}{45M[1-4M^2(90\pi-7\Omega_H^2)+16M^4(540\pi\Omega_H^2+7\Omega_H^4)+64M^6(270\pi\Omega_H^4+\Omega_H^6)]}  \label{equation 62}
\end{equation}
For $\Omega_H=10^3$ and $M=M_{\textrm{crit}}=0.0002425$ we find directly from (\ref{equation 62}),
\begin{equation}
-\left(\frac{\partial\Omega_H}{\partial T}\right)_J=1.4875\times10^5.
\label{equation 68}
\end{equation}

To calculate the right hand side of (\ref{eren2}) we employ the graphical results, as stated earlier. This yields, 
\begin{equation}
\frac{(J\alpha_2-J\alpha_1)}{(Jk_{T_2}-Jk_{T_1})}=\frac{(2.2\times10^{-5})}{(4.1\times 10^{-10})}=5.37\times10^4.
\label{equation 67}
\end{equation}

The disagreement between (\ref{equation 68}) and (\ref{equation 67}) spoils the possibility of a second order phase transition in black hole thermodynamics. In the following section we shall argue that this phase transition is very special and usually found in glass forming liquids when they are supercooled to form glasses. 

\subsection{Glassy phase transition and Prigogine-Defay ratio for Kerr black hole}
When some liquids are supercooled they solidify without crystallisation and form glasses, most common of which is the silicate glass. Though the underlying properties of materials which favour a glassy phase transition have not been completely understood \cite{rawson}, it is known that the rate of cooling is an important factor. Metals and alloys which were previously known as non glass forming substances are now found to form glasses under very fast cooling ($10^5$ K/s)\cite{giessen}. This is almost $10^7$ times faster than the cooling rate to form normal glass. In spite of the fact that in the glassy phase transition the specific heat, expansion coefficient and isothermal compressibility suffer discontinuity at the critical temperature, one cannot identify this as second order phase transition for the following reason. The critical temperature and the width of the transformation region for the glassy transition depend on the cooling rate\cite{jackle}. As a consequence the second Ehrenfest's equation (\ref{eren2}) is violated for this type of phase change.

For a normal second order phase transition, first order derivatives of Gibb's free energy are continuous at the phase transition point and hence one can equate the left hand sides of two Ehrenfest's equations to get the result $\Pi=1$, where $\Pi$, the Prigogine-Defay ratio \cite{jackle}, is defined by
\begin{eqnarray}
\Pi=\frac{\Delta C_p\Delta k}{TV(\Delta \alpha)^2}
\label{equation 3b}
\end{eqnarray}
In the case of black hole phase transition, from (\ref{equation 58a}) and (\ref{equation 68}) we see that left hand sides of (\ref{eren1}) and (\ref{eren2}) are also identical. Since in the previous subsection we discussed about the violation of second Ehrenfest's equation, we rectify it by putting a correction term ($\Pi$) in the right hand side of (\ref{eren2}) to get
\begin{eqnarray}
-\left(\frac{\partial\Omega_{H}}{\partial T}\right)_{J}=\Pi\frac{\alpha_{2}-\alpha_{1}}{k_{T_2}-k_{T_1}}
\label{equation 69z}
\end{eqnarray}   
Equating (\ref{equation 69z}) and (\ref{equation 58a}) we get the Prigogine-Defay ratio (which is the black hole analogue of (\ref{equation 3b}))   
\begin{equation}
\Pi=\frac{\Delta C_{\Omega_H}\Delta k_{T}}{TJ(\Delta\alpha)^2}.
\label{equation 70}
\end{equation}
For the Kerr black hole an explicit calculation gives $\Pi=\frac{0.0222\times 4.1\times10^{-10}}{0.006756\times{0.000022}^2}=2.78$. This is strikingly well placed in the known limit $2\leq\Pi\leq 5$ for glassy phase transitions in liquids and polymers \cite{gupta}.

Our studies further reveal that for values of  $\Omega_H$ ($\geq 10^2$), the critical point ($M_{\textrm{crit}}$ and $\tilde T_{\textrm{crit}}$) is shifted but the value of PD ratio remains same. For small values of $\Omega_H$ ($<10$), the change in $Jk_{T}$ between two phases is extremely small and it is very difficult to check the validity of Ehrenfest's equations. When $\Omega_H=10$ both of the Ehrenfest's equations are violated and only further studies can reveal the peculiarity of this point. We thus infer that the glassy phase transition occurs for $\Omega_H \gtrsim 10$.  

One may wonder whether the results are dependent on the normalisation of the logarithmic term appearing in (\ref{equation 21}). While it is difficult to answer this question in complete generality, nevertheless, we provide some arguments that establish the robustness of our conclusions. The first point to note is that, contrary to other black holes (like Schwarzschild and Reissner-Nordstrom etc) where different approaches yield different normalisations \cite{OTHER}, explicit results for the Kerr metric are only given in \cite{Modak}. However that does not prevent us from taking other normalisations and testing the inferences. Specifically, if the normalisation of the logarithmic term in (\ref{equation 21}) is taken as $\frac{1}{2}$ (instead of $\frac{1}{90}$) we again get a PD ratio close to 3, signifying a similar type of glassy phase transition. Likewise, if we consider other normalisations like $\frac{3}{2}$ (which, contrary to the earlier $\frac{1}{90}$ or $\frac{1}{2}$, is greater than one) or 1, our basic result remains unaffected. For details, see appendix 2. On the other hand, a negative normalisation of the logarithmic correction term modifies the corrected Hawking temperature in such a way that it is found to be negative in the entire non-extremal region which is clearly unphysical. So our conclusion is strictly valid for a positive normalisation of the correction term. 


\section{Conclusions}
In spite of considerable research in black hole physics, issues related to thermodynamical stability are often overlooked. This is because in most of the cases black holes are studied within the semi-classical regime where it is always thermodynamically unstable. This implies that the possibility of phase transition and stability can be studied by going beyond the semi-classical approximation. Also, it is necessary to have the work term ($PdV$-type) in the law of black hole thermodynamics to properly identify and classify a phase transition. Both these criteria are satisfied by considering the corrected expressions for the thermodynamic variables in the Kerr black hole. These expressions were taken from an earlier work done by two of us \cite{Modak}. This reference, however, did not consider at all the issues related to phase transition.

Here we showed that inclusion of correction beyond the semi-classical approximation makes the Kerr black hole stable via a phase transition. We discovered that this phase transition cannot be classified either as first or second order. This type of phase transition has a close analogy with liquid to glass transition and is known as a glassy phase transition. In particular, the Prigogine-Defay ratio found for the Kerr black hole fits within the bound obtained from experimental results for a glassy phase transition. Furthermore, we showed the robustness of our result by choosing three more arbitrary positive coefficients of the correction (logarithmic) term to the semi-classical entropy for the Kerr black hole. In all these cases a glassy phase transition occurred with almost the same PD ratio. This ratio, close to 3, was well within the bounds (2 to 5) quoted in the literature \cite{gupta} for glassy phase transitions in liquids and polymers. We feel that this work can open up a new field of research where the deep connection between black hole physics and thermodynamics (statistical mechanics) can be further analyzed.

\section*{Appendix 1}
\subsection*{Derivation of Ehrenfest's equations for Kerr black hole}
Using the first law of black hole thermodynamics (\ref{equation 3a}) and the definition of Gibb's function (\ref{equation 19a}), the differential form of Gibbs free energy is written as,
\begin{eqnarray}
dG=-Jd\Omega_H-SdT\label{dg}
\end{eqnarray}
From this equation, black hole entropy and the angular momentum can be written as the derivatives of the Gibbs free energy as,
\begin{eqnarray}
&&S=-\left(\frac{\partial G}{\partial T}\right)_{\Omega_H}\\
&&J= -\left(\frac{\partial G}{\partial \Omega_H}\right)_{T}
\end{eqnarray}
Since, by definition, the Gibbs free energy (\ref{equation 19a}) is a state function, $dG$ is an exact differential and hence we get the following Maxwell relation from (\ref{dg})
\begin{eqnarray}
\left(\frac{\partial J}{\partial T}\right)_{\Omega_H}=\left(\frac{\partial S}{\partial \Omega_H}\right)_{T}\label{maxwellrelation}
\end{eqnarray}
In a second order phase transition Gibbs free energy and its first order derivatives {\it i.e.} entropy and angular momentum, are all continuous. So at a phase transition point, characterized by some temperature $T$ and an angular velocity $\Omega_H$
\begin{eqnarray}
&&G_1=G_2\label{freeenergyequality}\\
&&S_1=S_2 \label{entropyequality}\\
&&J_1=J_2\label{angularmomentumequality}
\end{eqnarray}
where the subscripts 1 and 2 denote the values of the different physical quantities in the two phases.

Let us now consider the equality (\ref{entropyequality}). If the temperature and angular velocity are increased infinitesimally to $T+dT$ and $\Omega_H+d\Omega_H$ then
\begin{eqnarray}
S_1+dS_1=S_2+dS_2\label{2ndentropyequality}
\end{eqnarray}
From (\ref{entropyequality}) and (\ref{2ndentropyequality}) we find
\begin{eqnarray}
dS_1=dS_2\label{entropychange}
\end{eqnarray}
Taking $S$ as a function of $T$ and $\Omega_H$
\begin{eqnarray}
S=S(T,\Omega_H)
\end{eqnarray}
we write the infinitesimal change in entropy as
\begin{eqnarray}
dS=\left(\frac{\partial S}{\partial T}\right)_{\Omega_H}dT+\left(\frac{\partial S}{\partial \Omega_H}\right)_{T}d\Omega_H
\end{eqnarray}
Using (\ref{maxwellrelation}), the above equation takes the form,
\begin{eqnarray}
dS&=\frac{C_{\Omega_H}}{T}dT+J\alpha d\Omega_H
\end{eqnarray}
where 
\begin{equation}
\alpha=\frac{1}{J}\left(\frac{\partial J}{\partial T}\right)_{\Omega_H} 
\label{volexp}
\end{equation}
is the coefficient of change in angular momentum and 
\begin{equation}
C_{\Omega_H}=T\left(\frac{\partial S}{\partial T}\right)_{\Omega_H}. 
\label{corsp}
\end{equation}
Since $J$ is same in both phases,
\begin{eqnarray}
&&dS_1=\frac{C_{\Omega_{H_1}}}{T}dT+J\alpha_1 d\Omega_H \label{ds1}\\
&&dS_2=\frac{C_{\Omega_{H_2}}}{T}dT+J\alpha_2 d\Omega_H \label{ds2}
\end{eqnarray} 
Now use of the condition (\ref{entropychange}) requires the equality of (\ref{ds1}) and (\ref{ds2}),
\begin{eqnarray}
\frac{C_{\Omega_{H_1}}}{T}dT+J\alpha_1 d\Omega_H=\frac{C_{\Omega_{H_2}}}{T}dT+J\alpha_2 d\Omega_H.
\end{eqnarray} 
This gives the first Ehrenfest's equation  
\begin{eqnarray}
-\left(\frac{\partial\Omega_H}{\partial T}\right)_{S}=\frac{C_{\Omega_{H_2}}-C_{\Omega{_{H_1}}}}{TJ(\alpha_2-\alpha_1)}
\label{eren1}
\end{eqnarray} 
We now take the constancy of angular momentum (\ref{angularmomentumequality}) which, under infinitesimal change of temperature and angular velocity, gives
\begin{eqnarray}
dJ_1=dJ_2.\label{angularmomentumchange}
\end{eqnarray}
Taking angular momentum as a function of $T$ and $\Omega_H$,
\begin{eqnarray}
J=J(T,\Omega_H)
\end{eqnarray}
we get the differential relation
\begin{eqnarray}
dJ&=&\left(\frac{\partial J}{\partial T}\right)_{\Omega_H}dT+\left(\frac{\partial J}{\partial \Omega_H}\right)_Td\Omega_H\\
&=&J\alpha dT-Jk_Td\Omega_H
\end{eqnarray}
where 
\begin{equation}
k_T=\frac{1}{J}\left(\frac{\partial J}{\partial {\Omega_H}}\right)_T
\label{compr}
\end{equation}
is the analog of compressibility. Since $dJ$ is same for the two phases, we get the second Ehrenfest's equation by mimicking the previous steps used to derive (\ref{eren1})
\begin{eqnarray}
-\left(\frac{\partial\Omega_{H}}{\partial T}\right)_{J}=\frac{\alpha_{2}-\alpha_{1}}{k_{T_2}-k_{T_1}}.
\label{eren2}
\end{eqnarray} 
The two Ehrenfest's equations (\ref{eren1}) and (\ref{eren2}) which must be satisfied for a second order phase transition play the same role as the Clapeyron's equation for the first order phase transition.

\section*{Appendix 2}
\subsection*{Summary of results for distinct normalisations of the logarithmic term in (\ref{equation 21})}
Here we summarise our findings for different normalisations of the logarithmic term in (\ref{equation 21}). For instance, if the normalisation is taken as $\frac{1}{2}$, the corrected entropy and temperature are given by, 
\begin{eqnarray}
\tilde S=\frac{A}{4\hbar}+\frac{1}{2}\log{\frac{A}{4\hbar}}+{\cal O}(\hbar), 
\label{diffs1}
\end{eqnarray}
and
\begin{eqnarray}
\tilde T=T\left(1+\frac{\sqrt\hbar}{4\pi Mr_+}+{\cal O}(\hbar^2)\right)^{-1}.
\label{difft1}
\end{eqnarray}    
After carrying out the analysis with these corrected expressions one would find the following results for two Ehrenfest's equations, which are the analogues of (\ref{equation 58a}) and (\ref{equation 58}), obtained with the earlier normalisation ($\frac{1}{90}$),
\begin{eqnarray}
-\left(\frac{\partial\Omega_H}{\partial T}\right)_{S}=6.6934\times10^{6}\nonumber\\
\frac{C_{\Omega_{H_2}}-C_{\Omega{_{H_1}}}}{TJ(\alpha_2-\alpha_1)}=7.0109\times 10^{6}.
\label{diff1eren1}
\end{eqnarray}  
and 
\begin{eqnarray}
-\left(\frac{\partial\Omega_{H}}{\partial T}\right)_{J}=6.68077\times10^{6}\nonumber\\
\frac{\alpha_{2}-\alpha_{1}}{k_{T_2}-k_{T_1}}=2.2500\times 10^{6}.
\label{diff1eren2}
\end{eqnarray} 
The above pair of equations (\ref{diff1eren2}) are the analogues of (\ref{equation 68}) and (\ref{equation 67}). This clearly shows that while the first Ehrenfest's equation is closely satisfied the second one is violated in such a way that the PD ratio (\ref{equation 70}) following from (\ref{diff1eren2}) becomes $\Pi=\frac{6.68077\times10^{6}}{2.2500\times10^{6}}=2.96$.

Similarly, taking the normalisation as $\frac{3}{2}$ of the logarithmic term in (\ref{equation 21}), we get
\begin{eqnarray}
\tilde S=\frac{A}{4\hbar}+\frac{3}{2}\log{\frac{A}{4\hbar}}+{\cal O}(\hbar), 
\label{diffs2}
\end{eqnarray}
\begin{eqnarray}
\tilde T=T\left(1+\frac{3\sqrt\hbar}{4\pi Mr_+}+{\cal O}(\hbar^2)\right)^{-1}.
\label{difft2}
\end{eqnarray}   
Correspondingly, the results for the two Ehrenfest's equations are given by
\begin{eqnarray}
-\left(\frac{\partial\Omega_H}{\partial T}\right)_{S}=2.00802\times10^{7}\nonumber\\
\frac{C_{\Omega_{H_2}}-C_{\Omega{_{H_1}}}}{TJ(\alpha_2-\alpha_1)}=2.14084\times 10^{7}.
\label{diff2eren1}
\end{eqnarray}  
and 
\begin{eqnarray}
-\left(\frac{\partial\Omega_{H}}{\partial T}\right)_{J}=2.00423\times10^{7}\nonumber\\
\frac{\alpha_{2}-\alpha_{1}}{k_{T_2}-k_{T_1}}=6.82923\times 10^{6}.
\label{diff2eren2}
\end{eqnarray} 
These results finally give a PD ratio $\Pi=2.93$ for the above choice of the normalisation. 

Finally, if we set the normalisation as unity, the corrected temperature and entropy are given by,
\begin{eqnarray}
\tilde S=\frac{A}{4\hbar}+\log{\frac{A}{4\hbar}}+{\cal O}(\hbar), 
\label{diffs3}
\end{eqnarray}
\begin{eqnarray}
\tilde T=T\left(1+\frac{\sqrt\hbar}{2\pi Mr_+}+{\cal O}(\hbar^2)\right)^{-1},
\label{difft3}
\end{eqnarray}   
leading to the Ehrenfest's equations,
\begin{eqnarray}
-\left(\frac{\partial\Omega_H}{\partial T}\right)_{S}=1.33868\times10^{7}\nonumber\\
\frac{C_{\Omega_{H_2}}-C_{\Omega{_{H_1}}}}{TJ(\alpha_2-\alpha_1)}=1.36735\times 10^{7}.
\label{diff3eren1}
\end{eqnarray}  
and 
\begin{eqnarray}
-\left(\frac{\partial\Omega_{H}}{\partial T}\right)_{J}=1.33615\times10^{7}\nonumber\\
\frac{\alpha_{2}-\alpha_{1}}{k_{T_2}-k_{T_1}}=4.63415\times 10^{6}.
\label{diff3eren2}
\end{eqnarray} 
Once again one finds a PD ratio $\Pi=2.88$ which closely matches with other cases.
\section*{Acknowledgement} 
One of the authors (S.K.M) thanks the Council of Scientific and Industrial Research (C.S.I.R), Government of India, for financial support.

\end{document}